\documentclass[12pt]{article}

\usepackage{latexsym}   
\input tcilatex
\QQQ{Language}{
American English
}

\begin{document}

\title{Nonlinear Hodge equations in vector bundles}
\author{Thomas H. Otway \\
\\
\textit{Departments of Mathematics and Physics,}\\
\textit{Yeshiva University, New York, New York 10033}}
\date{}
\maketitle

\begin{abstract}
A gauge-invariant form of the nonlinear Hodge equations is studied. 1991
MSC: 58E15 (Classical field theory)
\end{abstract}

\section{\textrm{Introduction}}

Under suitably harsh assumptions, many natural phenomena can be expressed as
solutions of the linear Hodge-Kodaira equations, in which stationary fields
appear as harmonic forms. If the drastic physical assumptions of the linear
theory are relaxed, then at first glance the Hodge-de Rham interpretation
appears to crumble before a bewildering variety of nonlinear variational
theories. Nonlinear Hodge theory, introduced nearly 30 years ago by L. M.
Sibner and R. J. Sibner \textbf{[SS1, SS2]}, can be viewed as an extension
of the unified geometric interpretation achieved for linear field equations
to the quasilinear case. (Specifically, a nonlinear ''mass density'' term is
introduced into the linear Hodge-Kodaira equations for differential forms on
a Riemannian manifold. If this mass density term is constant, then the
nonlinear Hodge equations reduce to the linear Hodge-Kodaira equations.)

Although many of the results of nonlinear Hodge theory extend to
differential forms of arbitrary degree (see, \textit{e.g.,} [\textbf{Si1}]
and [\textbf{SS4}]), 1-forms occupy a special place in that a 1-form which
is closed under exterior differentiation can be interpreted as the gradient
of a 0-form (or, in physical terms, as the field of a scalar potential).
This interpretation is exploited in [\textbf{SS1}]-[\textbf{SS3}]. A 2-form
which is closed under \textit{covariant} exterior differentiation can, in
certain circumstances, be interpreted as the curvature of a connection
1-form (or, in physical terms, as the field of a vector potential). This
interpretation leads to enriched geometry, as the relevant bundle need no
longer be the cotangent bundle of a manifold, as in conventional nonlinear
Hodge theory, but can be a bundle with curvature.

This is the motivation for generalizing aspects of nonlinear Hodge theory to
sections of a vector bundle having nonabelian structure group. This
extension adds extra nonlinearities and nontrivial gauge invariance to the
equations of\textbf{\ [SS1, SS2]}. From a variational point of view, its aim
is to draw certain nonquadratic energies of Yang-Mills type into the family
of variational integrals amenable to a nonlinear Hodge-de Rham
interpretation. From a geometric point of view, the generalized equations
obtained bear the same relation to harmonic curvature on a bundle that the
nonlinear Hodge equations bear to harmonic forms on a manifold, leading to a
quasilinear generalization of harmonic curvature. From a purely analytic
point of view, the extra nonlinearities of the bundle-valued equations yield
new insights into the form-valued equations. For example, these ideas lead
to a weakening of the conventional ''irrotationality'' assumption for
stationary nonlinear Hodge flow on a manifold [\textbf{O4}].

We introduced the main features of this generalized nonlinear Hodge theory
in two previous papers [\textbf{O2}], [\textbf{O3}]. Here we consider a
variety of technical points, some of which were ignored in earlier work,
others only sketched, and still others treated inadequately. As a
consequence we obtain revised proofs of some of the results in [\textbf{O2}%
], [\textbf{O3}].

We note that energy functionals which are nonquadratic in the bundle
curvature have already appeared in the physics literature in the context of
individual models - for example, the well known model introduced by
Tchrakian for higher-dimensional gauge theories [\textbf{T}].

In the sequel we denote by \textit{C} generic positive constants which
generally depend on dimension and which may change in value from line to
line.

\section{A nonlinear Hodge theory for 2-forms}

Let $M$ be a finite, oriented, \textit{n}-dimensional Riemannian manifold
and $X$ a vector bundle over $M$ having compact structure group $G\subset
SO(m).$ Let $A\in \Gamma \left( M,ad\,X\otimes T^{*}M\right) $ be a
connection 1-form with curvature 2-form $F_A,$ where 
\[
F_A=dA+\frac 12\left[ A,\,A\right] =dA+A\wedge A 
\]
and [\ ,\ ] is the Lie bracket of the Lie algebra $\Im $ associated to $G$.
Geometrically, $\Im $ is the fiber of the adjoint bundle $ad\,X.$ Here $%
d:\Lambda ^p\rightarrow \Lambda ^{p+1}$ is the flat exterior derivative and $%
\wedge ,$ the wedge product on differential forms. Sections of the
automorphism bundle $Aut\,X$ are called \textit{gauge transformations}.
These act tensorially on $F_A$ but affinely on $A$, a fact which leads to
certain analytic difficulties. For details of this geometric construction
see, \textit{e.g.,} [\textbf{MM}].

We consider a stored energy functional of the form 
\begin{equation}
E=\frac 12\int_M\left( \int_0^Q\rho (s)ds\right) dM,
\end{equation}
where $Q=|F_A|^2=\left\langle F_A,F_A\right\rangle $ is an inner product on
the fibers of the bundle $ad\,X\otimes \Lambda ^2\left( T\;^{*}M\right) $
(the inner product on $ad\,X$ being induced by the normalized trace inner
product on $SO(m)$ and that on $\Lambda ^2\left( T\;^{*}M\right) ,$ by the
exterior product $*\left( F_A\wedge *F_A\right) $, where$\,*:\Lambda
^p\rightarrow \Lambda ^{n-p}$ is the Hodge involution); $\rho :\mathbf{R}%
^{+}\rightarrow \mathbf{R}^{+}$ is a \textit{C}$^1$ function satisfying

\begin{equation}
K^{-1}(Q+k)^q\leq \rho (Q)+2Q\rho ^{\prime }(Q)\leq K(Q+k)^q
\end{equation}
for some positive constant $K$ and nonnegative constants $k$, $q$.

The functional (1) is a generalization of the \textit{nonlinear Hodge energy}
introduced in \textbf{[SS2]} for $X=T\;^{*}M$ and $Q=|\omega |^2,$ where $%
\omega \in \Gamma \left( M,\Lambda ^p\left( T^{*}M\right) \right) .$ (See
also \textbf{[U1]}, page 221.) Critical points of (1) with respect to an
admissible cohomology class of closed $p$-forms satisfy the \textit{%
nonlinear Hodge equations} 
\begin{equation}
\delta \left( \rho (Q)\omega \right) =0,
\end{equation}
\begin{equation}
d\omega =0,
\end{equation}
which were introduced and extensively studied by L. M. and R. J. Sibner [%
\textbf{SS1}]-[\textbf{SS4}]. Choosing, for $X=T\;^{*}M,$ $p=1,$ and $\omega
=d\varphi ,\rho (Q)$ to be

\[
\rho (Q)=\left( 1-\frac{\gamma -1}2Q\right) ^{1/(\gamma -1)}, 
\]
where $\gamma >1$ is a constant, provides (1) with an interpretation as the
energy functional for the stationary, polytropic flow of a compressible
fluid having adiabatic constant $\gamma .$ The scalar $\varphi $ is a
possibly multivalued potential for the velocity field $\omega .$ The
multivalued nature of $\varphi $ describes circulation of the flow, \textit{%
e.g.,} about an obstacle with handles. In this case inequality (2) with $q=0$
is a condition for subsonic flow. $Q_{crit}=2/(\gamma +1)$ is the squared
speed at the transition from subsonic to supersonic flow. The Euler-Lagrange
equations for variations of (1) with $p=1$ yield the continuity equations
for the flow in Eulerian coordinates [\textbf{SS1}]. Analogous
interpretations can be given to topics in elasticity and thermodynamics,
including nonrigid-body rotation and capillarity. Applications to magnetic
materials and minimal surfaces are given in [\textbf{O3}] and [\textbf{SS2}%
], respectively.

In order to extend the variational problem to sections of a vector bundle,
we form an admissible class of \textit{connections} by choosing a smooth
base connection $D$ in the space of connections compatible with $G$ and
considering the class of connections $D+A,$ where $A$ is a section of $%
ad\,X\otimes T^{*}M$ which lies in the largest Sobolev space for which the
energy $E$ is finite; details are given in [\textbf{U3}] for the case $\rho
\equiv 1$. We take variations by computing $(d/dt)(F_{D+tA})$ at the origin
of $t.$ Using the fact that for any smooth section $\sigma $ we have

\[
F_{D+tA}(\sigma )=\left( F+tDA+t^2A\wedge A\right) (\sigma ), 
\]
we obtain

\begin{equation}
\delta E=\frac 12\int_M\rho (Q)\delta Q\;dM=\frac 12\int_M\rho (Q)\frac
d{dt}_{|t=0}|F+tDA+t^2A\wedge A|^2\;dM.
\end{equation}
Notice that $D=d+[A,\;].$ Letting $t=0,$ the right-hand side of (5) can be
written

\begin{equation}
\int_M\rho (Q)\,\langle DA,\,F_A\rangle \;dM=\int_M\langle DA,\,\rho
(Q)F_A\rangle \;dM.
\end{equation}
We assume that either $\partial M=0$ or, if not, that $F_A$ satisfies a
''Neumann'' boundary condition of the form 
\begin{equation}
i^{*}(*F)=0
\end{equation}
on $\partial M,$ where $i^{*}$ is the pull-back under inclusion of the
boundary of $M$ in $M$. This is equivalent in local coordinates to
prescribing zero boundary data for $F$ in a direction normal to $\partial M;$
see, \textit{e.g.,} [\textbf{Ma}] for details in the case $\rho \equiv 1.$

Set $\delta E=0$ equal to zero. Then (5) and (6) imply 
\[
0=\int_M\left\langle DA,\,\rho (Q)F_A\right\rangle \;dM= 
\]
\[
\int_Md\left( A\wedge *\,\rho (Q)F_A\right) +\int_M\left\langle
A,\,D^{*}\left( \rho (Q)F_A\right) \right\rangle \;dM 
\]
\[
=\int_{\partial M}A_\vartheta \wedge \left( \,\rho (Q)F_A\right)
_N\,+\int_M\left\langle A,\,D^{*}\left( \rho (Q)F_A\right) \right\rangle
\;dM, 
\]
where $D^{*}$ denotes the formal adjoint of the exterior covariant
derivative $D;$ $\vartheta $ denotes tangential component on the boundary
and $N$, the normal component there. Condition (7) implies Euler-Lagrange
equations of the form 
\begin{equation}
D^{*}\left( \rho (Q)F\right) =0.
\end{equation}
Because $F$ is a curvature 2-form, it satisfies an additional condition 
\begin{equation}
DF=0.
\end{equation}
(This is the second Bianchi identity.) This paper is concerned with analytic
properties of the system (8), (9).

If we write these equations as a system of equations for Lie-algebra-valued
forms, they can be written 
\begin{equation}
\delta \left( \rho (Q)F_A\right) =-*\left[ A,*\rho (Q)F_A\right] ,
\end{equation}
\begin{equation}
dF_A=-\left[ A,F_A\right] .
\end{equation}
Here $\delta :\Lambda ^p\rightarrow \Lambda ^{p-1}$ is the adjoint of the
exterior derivative $d$. If $G$ is abelian, then the Lie brackets in (10),
(11) vanish and eqs. (8), (9) reduce to the system 
\[
\delta \left( \rho \left( Q(F)\right) F\right) =dF=0, 
\]
which are the nonlinear Hodge equations for the 2-form $F$ in a local
trivialization of $X$. If in addition $\rho \equiv 1,$ then we obtain the
Hodge-Kodaira equations for 2-forms. If $G$ is nonabelian and $\rho \equiv
1, $ then eqs. (8) reduce to the \textit{Yang-Mills equations,} the
equations for the classical limit of quantum fields.

Equations (10) have certain formal similarities to the continuity equation
for a velocity field, having components $v^\alpha ,$ of a stationary,
polytropic, compressible fluid on a Riemannian manifold $M$ possessing a $%
C^1 $ metric tensor $g_{\alpha \beta }$ and affine connection $\Gamma
_{\gamma \beta }^\alpha ,$ where $\alpha ,\beta =1,\,...,\,n$, that is, 
\begin{equation}
\partial _\alpha \left( \rho (Q)v^\alpha \right) +\rho (Q)v^\alpha \Gamma
_{\alpha \beta }^\beta =0.
\end{equation}
Here $\partial _\alpha =\partial /\partial x^\alpha $ for $%
x=(x^1,\,...,x^n)\in M$ and $Q=g^{\alpha \beta }v_\alpha v_\beta $. If the
flow is parallel on $M,$ then 
\begin{equation}
\partial _\beta v^\alpha +v^\gamma \Gamma _{\gamma \beta }^\alpha =0,
\end{equation}
which is to say that the covariant derivative of $v$ vanishes with respect
to the connection $\Gamma _{\gamma \beta }^\alpha .$ Equation (13) is thus
the geometric analogue for parallel 1-tensors on a manifold of the Bianchi
identity (9) for curvature 2-forms (on a vector bundle). If $g\equiv \det
\left( g^{\alpha \beta }\right) $ is a $C^1$ function of $x,$ then we can
write 
\[
\Gamma _{\alpha \beta }^\beta =\frac 1{\sqrt{g}}\frac \partial {\partial
x^\alpha }\sqrt{g}. 
\]
On a Riemannian manifold the operator $\delta $ explicitly involves the
metric: 
\[
\delta _M(\vartheta )=-\frac 1{\sqrt{g}}\frac \partial {\partial x^\alpha
}\left( \sqrt{g}\vartheta _\alpha \right) . 
\]
Applying the product rule to $\delta _M(\vartheta )$ with $\vartheta =\rho
(Q)v,$ we conclude that on a differentiable Riemannian manifold eq. (3) is
exactly dual to (12), as the former equation results from replacing the
tangent bundle in the latter equation by the cotangent bundle. (This
relation between (3) and (12) was introduced in [\textbf{SS1}].) But eq. (4)
asserts that the flow is \textit{irrotational:} there is no circulation
about any curve homologous to zero. In euclidean space every parallel flow
is irrotational, but this is not true in general (the simplest example being
flow along a great circle of a sphere), so on a manifold condition (4)
differs somewhat from condition (13) $-$ \textit{c.f.} [\textbf{O4}].

In the case of abelian $G$, for given $\lambda \in \overline{Ker\,d}$ an
admissible class is defined by the set of $\omega \in \overline{Ker\,d\text{ 
}}$ for which $\omega -\lambda \in \overline{\func{Im}\,d}.$ This condition
prescribes a cohomology class of admissible forms, leading to a very
complete existence theory for 1-forms \textbf{[SS2]}. This theory extends to 
\textit{p-}forms on a compact Riemannian manifold \textbf{[Si1]}, but
appears to fail for nonabelian $G$. Another analytic difficulty is that, in
distinction to the conventional Yang-Mills equations, eqs. (8), (9) cannot
be written as a diagonal elliptic system even in a good gauge. Two obvious
consequences are that the technique used in [\textbf{Ma}] to solve
boundary-value problems for the 4-dimensional Yang-Mills equations will not
work for (8), (9) and that H\"{o}lder continuity for solutions of (8), (9)
does not automatically imply any higher regularity.

\section{The smoothness of solutions}

In this section we derive regularity and gauge-improvement results for
solutions of the system (8), (9).

The following is a revision of Theorem 1.1 of [\textbf{O2}]. (The
unnecessary restriction that $Q$ be positive is removed; an estimate for
nonlinear Hodge fields [\textbf{O4}] is used to obtain a bound on the bundle
curvature; the role of the exponential gauge in the proof is clarified; a
gauge-invariant Campanato estimate is constructed.)

\begin{theorem}
Let the pair $(A,F_A)$ weakly satisfy eqs. (8), (9) in a bounded, open,
type-A domain $\Omega \subset \mathbf{R}^n.$ Let $\rho $ satisfy condition
(2) with $q=0$. Suppose that $F_A\in L^s(\Omega )$ for some $s>n/2.$ Then $A$
is equivalent via a continuous gauge transformation to a connection $%
\widetilde{A}$ such that $F_{\widetilde{A}}$ is H\"{o}lder continuous in $%
\Omega .$
\end{theorem}

\textbf{Remarks.}\textit{\ i) }In the following proof we take the $L^n$-norm
of $A$ to be small on a sufficiently small ball (in the sense of an \textit{n%
}-disc). In a part of the proof sketched in an appendix to this paper we
additionally require the $L^{n/2}$-norm of $F$ to be small on a small ball.
Neither asumption need be stated explicitly in Theorem 1. If $F\in L^s$ for $%
s>n/2$, the small-$L^{n/2}$-norm assumption for $F$ follows from standard
arguments on $\mathbf{R}^n$ (and from Lemma 3.4 of [\textbf{U3}] on a
Riemannian manifold). The corresponding assumption for $A$ follows from from
Theorem 1.3\textit{\ ii)} of [\textbf{U3}] and the Sobolev Theorem.

\textit{ii)} We exploit the boundedness of $\Omega $ to study eqs. (8), (9)
in a small $n$-disc $B$ and employ a covering argument at the end. This
allows us to trivialize $X$ locally and understand the notion of weak
solution in the sense of [\textbf{Si1}], eq. (1.2b). For abelian $G$, a 
\textit{weak solution} of (8), (9) is any curvature 2-form $F_A$ for which $%
\rho (Q)F_A$ is orthogonal in $L^2$ to the space of $d$-closed 2-forms $%
d\zeta \in L^2(B)$ such that $\zeta \in \Lambda ^1$ has vanishing tangential
data on $\partial B.$ For nonabelian $G$, an obvious extension of (1.2b) to
inhomogeneous equations allows us to define a weak solution of (8), (9) by
the equation 
\begin{equation}
\int_B\left\langle d\zeta ,\rho (Q)F_A\right\rangle *1=-\int_B\left\langle
\zeta ,*\left[ A,*\rho (Q)F_A\right] \right\rangle *1,
\end{equation}
where $F$ is a curvature 2-form. Our general understanding of weak solutions
to gauge-invariant systems is derived from [\textbf{U3}].

\textit{iii)} For a definition of \textit{type-A domain} see, \textit{e.g.,}
[\textbf{Gi}]. As an example, any Lipschitz domain is type-\textit{A}.

\textit{iv)} It is easy to show the existence of weak solutions to (8), (9)
by topological arguments, provided that $\rho $ is chosen so that the energy
functional is Palais-Smale. An example is given in Corollary 1.2 of [\textbf{%
O2}].

\textit{v)} Theorem 1 cannot be improved (for $q=0$) without improving the
existing regularity theory [\textbf{U3}] for the case $\rho =1.$

The proof of Theorem 1 strongly uses the properties of the exponential gauge
in a euclidean $n$-disc $B$ centered at the origin of coordinates in $%
\mathbf{R}^n$, namely, that in such a gauge 
\[
A(0)=0 
\]
and $\forall x\in B$%
\[
\left| A(x)\right| \leq \frac 12\left| x\right| \cdot \sup_{\left| y\right|
\leq \left| x\right| }\left| F(y)\right| 
\]
(see [\textbf{U2}], Sec. 2).

We also require the following mean-value formula of L. M. Sibner, originally
stated for differential forms on a Riemannian manifold, which extends
immediately to the case of Lie-algebra-valued sections:

\begin{lemma}
(L. M. Sibner[\textbf{Si1}], Lemma 1.1). Let 
\[
G^J(x,\omega )=\sqrt{g}g^{IJ}\rho \left( Q(\omega )\right) \omega _I, 
\]
where $g^{ij}$ is the metric tensor on a compact Riemannian manifold $M;x\in
M;g=\det \left( g^{ij}\right) ;I,J$ are multi-indices; $\omega \in \Gamma
\left( M,\Lambda ^p(T^{*}M)\right) .$ Let $\rho $ satisfy condition (2) with 
$q=0$. Then for $\xi ,\eta \in M$ and $\mu ,\tau \in \Gamma \left( M,\Lambda
^p(T^{*}M)\right) ,$%
\[
G^J(\xi ,\mu )-G^J(\eta ,\tau )=\alpha ^{IJ}\left( \mu _I-\tau _I\right)
+\beta _i^J\left( \xi ^i-\eta ^i\right) , 
\]
where $\alpha ^{IJ}$ is a positive-definite matrix and 
\[
\left| \beta _i^J\right| \leq C\left( \left| \mu (x)\right| +\left| \tau
(x)\right| \right) . 
\]
\end{lemma}

Finally, we need an \textit{a priori} estimate for smooth solutions:

\begin{lemma}
Let the pair $(A,F_A)$ smoothly satisfy eqs. (8), (9) and condition (2) on
an open, bounded domain $\Omega \subset \mathbf{R}^n.$ Then the scalar $%
Q=|F|^2$ satisfies the inequality 
\begin{equation}
L(Q)+C(Q+k)^q\left( \left| \nabla A\right| +\left| A\right| ^2\right) Q\geq
0,
\end{equation}
where $L$ is a divergence-form operator which is uniformly elliptic for $k>0$%
.
\end{lemma}

\textit{Proof of Lemma 3.} The proof is nearly identical to the proof of [%
\textbf{O4}], Theorem 1, taking (in the notation of [\textbf{O4}]) $u=A$ and 
$\omega =F.$ However, eq. (10) contains an inhomogeneous term on the right
that is absent in eq. (1) of [\textbf{O4}]. This inhomogeneous term arises
as the second term on the right in the equation 
\[
\left\langle F,\Delta \left( \rho (Q)F\right) \right\rangle =-\left\langle
F,\delta d\left( \rho (Q)F\right) \right\rangle +\left\langle F,d*\left[
A,*\rho (Q)F\right] \right\rangle 
\]
(\textit{c.f.} [\textbf{O4}], eq. 14). This term can be estimated by Young's
inequality: 
\[
\left| F\right| \left| d*\left[ A,*\rho (Q)F\right] \right| \leq C\left(
\left| \nabla A\right| |\rho |Q+|F||A|\left| \bigtriangledown (\rho
F)\right| \right) 
\]
\[
\leq C(Q+k)^q\left( \left| \nabla A\right| Q+\varepsilon \left| \nabla
F\right| ^2+C(\varepsilon )|A|^2Q\right) 
\]
(\textit{c.f.} [\textbf{O4}], (18)). The remainder of the derivation of (15)
is exactly analogous to the derivation of inequality (7) in [\textbf{O4}],
provided the wedge product of differential forms on the right in eq. (2) of [%
\textbf{O4}] is replaced by a Lie bracket of Lie-algebra-valued sections.

Concerning the ellipticity of the operator $L$, define a function $h(Q)$
such that 
\[
h^{\prime }(Q)=\frac 12\rho (Q)+Q\rho ^{\prime }(Q) 
\]
and an operator $\widetilde{L}$ such that 
\[
\widetilde{L}\left( h(Q)\right) =\sum_{k,j}\partial _k\left\{ \left[ \delta
_{kj}-\frac{\rho ^{\prime }(Q)}{h^{\prime }(Q)}\left\langle \sigma
_kF,\sigma _jF\right\rangle \right] \partial _kh(Q)\right\} . 
\]
Here $\delta _{kj}$ is the kronecker delta and $\sigma _k$ is an
antisymmetrization operator equivalent to the operator $A_i(1)$ in eq. (1.9)
of [\textbf{U1}]. Then 
\[
\widetilde{L}\left( h(Q)\right) =\sum_{k,j}\partial _k\left\{ \left[ \frac
12\rho (Q)+Q\rho ^{\prime }(Q)-\rho ^{\prime }(Q)\left\langle \sigma
_kF,\sigma _jF\right\rangle \right] \partial _kQ\right\} \equiv L(Q). 
\]
The ellipticity of $\widetilde{L}$ under condition (2) with $k>0$,
established on p. 233 of [\textbf{U1}], implies the ellipticity of $L$ under
the same hypothesis. This completes the proof of Lemma 3.$\Box $

\textit{Proof of Theorem 1.} Because $F_A\in L^s(\Omega )$ for some $s>n/2$
there is a continuous gauge transformation in a small disc $B\subset \subset
\Omega $ to a \textit{Hodge gauge} in which the following conditions are
satisfied ([\textbf{U3}], Theorem 2.1):

\textit{a)} $\delta A=0;$

\textit{b)} $x\cdot A=0$ on $\partial B;$

\textit{c)} $\left\| A\right\| _{1,n/2}\leq C\left\| F\right\| _{n/2};$

\textit{d)} $\left\| A\right\| _{1,s}\leq C\left\| F\right\| _s$ for $%
n/2<s<n.$

Here $\left\| \cdot \right\| _{p,q}$ is the $H^{p,q}$-norm and $\left\|
\,\cdot \,\right\| _p$ is the $L^p$-norm on $B$. Condition \textit{a)} and
the Neumann condition \textit{b)} allow us to apply the Gaffney-G\aa rding
inequality [\textbf{Ga}] 
\[
\left\| \nabla A\right\| _2^2\leq C\left( \left\| dA\right\| _2^2+\left\|
\delta A\right\| _2^2+\left\| A\right\| _2^2\right) 
\]
in the form 
\begin{equation}
\left\| \nabla A\right\| _2^2\leq C\left\| dA\right\| _2^2
\end{equation}
(see the proof of [\textbf{U3}], Lemma 2.5), provided that we choose $B$ so
that $\left\| A\right\| _n$ is small.

Estimating the difference quotient of $F$ as in the proof of Lemma 3.1 of [%
\textbf{O2}], using properties \textit{a)-d),} inequality (16), and the fact
that exterior operators commute with the difference-quotient operator, we
find that $F\in H^{1,2}(\widetilde{B}),$ where $\widetilde{B}\subset \subset
B.$ For the reader's convenience this argument is briefly reviewed in an
appendix. At this point we would be able to apply Theorem 5.3.1 of [\textbf{%
Mo}], using Lemma 3 with $q=0$, provided we knew that $|F|^\tau $ was in $%
H^{1,2}(B^{\prime })$ for some $\tau >1$ and some $B^{\prime }\subset 
\widetilde{B.}$ That this condition is in fact satisfied can be seen be
writing inequality (15) in the weak form 
\[
\int_{\widetilde{B}}a^{ij}\partial _iQ\partial _j\zeta *1=2\int_{\widetilde{B%
}}a^{ij}u\frac{\partial u}{\partial x^i}\frac{\partial \zeta }{\partial x^j}%
d^nx\leq C\left\{ \int_{\widetilde{B}}\left( \left| \nabla A\right| +\left|
A\right| ^2\right) u^2\zeta d^nx\right\} , 
\]
where $u=|F|;$ the matrix $a^{ij}$ satisfies the ellipticity condition $%
m_1|\xi |^2\leq a^{ij}\xi _i\xi _j\leq m_2|\xi |^2$ for positive constants $%
m_1$ and $m_2$; $\zeta \in C_0^\infty (\widetilde{B})\cap \mathbf{R}/\mathbf{%
R}^{-}.$ Choose 
\[
\zeta =(u_k+\delta )^{2\tau -2}\eta ^2 
\]
for $\eta \in C_0^\infty (\widetilde{B}),\eta \geq 0,\delta >0,\tau >1.$ The
sequence $\left\{ u_k\right\} $ is chosen to be increasing and so that $%
\lim_{k\rightarrow \infty }u_k=u.$ We have 
\[
\int_{\widetilde{B}}a^{ij}u(u_k+\delta )^{2\tau -3}\partial _iu\,\partial
_j\left( u_k\right) \eta ^2*1\leq C\int_{\widetilde{B}}u\left| \nabla
u\right| (u_k+\delta )^{2\tau -2}\eta \left| \nabla \eta \right| *1 
\]
\[
+C\int_{\widetilde{B}}\left( \left| \nabla A\right| +\left| A\right|
^2\right) u^2(u_k+\delta )^{2\tau -2}\eta ^2*1\leq 
\]
\begin{equation}
\leq C\left\{ \int_{\widetilde{B}}(u+\delta )^{2\tau -1}\left| \nabla
u\right| *1+\int_{\widetilde{B}}\left( \left| \nabla A\right| +\left|
A\right| ^2\right) (u+\delta )^{2\tau }*1\right\} .
\end{equation}
The extreme right-hand side of inequality (17) can be bounded above by the
norms 
\[
\left\| u+\delta \right\| _{(2\tau -1)s_1}^{2\tau -1}\left\| \nabla
u\right\| _{s_2}+\left\| \left| \nabla A\right| +\left| A\right| ^2\right\|
_{s_3}\left( \int (u+\delta )^{2\tau s_4}*1\right) ^{1/s_4}. 
\]
These norms can be made finite for $\tau $ sufficiently close to 1. For $n>6$%
, choose $s_1=s_2=2,\;s_3=n/(n-4\tau ),\;s_4=n/(4\tau );$ then $s_3\leq s$
for $s>n/2$ if $\tau $ is close to 1. If $2<n\leq 6$ choose $s_1=s_2=2,$ $%
s_3=n/2\tau ,\;s_4=n/(n-2)\tau .$ The finite $H^{1,2}$-norm of $\left|
F\right| $ implies $u\in L^{2n/(n-2)}$ by the Sobolev Theorem and of course $%
\left\| A\right\| _{1,n/2\tau }\leq \left\| A\right\| _{1,n/2}<\infty .$ If $%
n=2,$ then the norms are finite for $s_1=s_2=s_3=s_4=2$ by the Sobolev
Theorem.

Using ellipticity, we obtain in place of (17) the estimate 
\[
\nu \int_{\widetilde{B}}\eta ^2\left| \nabla u^\tau \right| ^2d^nx\leq
C<\infty . 
\]
Letting $\eta =1$ on some smaller ball $B^{\prime }$ completely contained in 
$\widetilde{B}$ and concentric with it allows us to conclude that $|F|^\tau
\in H^{1,2}(B^{\prime })$ for some $\tau >1.$ Now we apply Theorem 5.3.1 of [%
\textbf{Mo}] to conclude that $|F|$ is bounded in $B^{\prime }.$

As gauge transformations act tensorially on $F,$ the curvature remains
bounded under continuous gauge transformations. In particular, at the origin
of coordinates in an exponential gauge eq. (10) becomes 
\[
\delta \left( \rho _0(Q)F_0\right) =\delta \left( \rho _0(Q)dA_0\right) =0, 
\]
where the subscript indicates that the result of the computation is being
evaluated at the origin of $B^{\prime }.$

Because X has been trivialized in $B^{\prime }$ we can compare $F$ to a
solution $d\varphi $ of the variational problem associated to the equation 
\begin{equation}
\int_{B^{\prime }}\left\langle d(A-\varphi ),\rho \left( Q(d\varphi )\right)
d\varphi \right\rangle *1=0.
\end{equation}
The 2-form $d\varphi $ exists as a weak $L^2$ solution by Proposition 4.3 of
[\textbf{S1}]; $d\varphi $ is H\"{o}lder continuous by Proposition 4.4 of [%
\textbf{S1}] (which is derived from [\textbf{U1}]). In particular, the test
function $d(A-\varphi )$ is admissible: 
\[
\left\| d\left( A-\varphi \right) \right\| _2\leq \left\| dA\right\|
_2+\left\| d\varphi \right\| _2, 
\]
where, in an exponential gauge, 
\[
\left\| dA\right\| _2\leq \left\| dA+\frac 12\left[ A,A\right] \right\|
_2+C\left\| A\right\| _2\leq C\left\| F\right\| _2<\infty . 
\]
Combining (18) with (14), we have 
\[
\int_{B^{\prime }}\left\langle d(A-\varphi ),\rho \left( Q(F)\right)
F_A-\rho \left( Q(d\varphi )\right) d\varphi \right\rangle
*1=\int_{B^{\prime }}\left\langle A-\varphi ,*\left[ A,*\rho (Q)F_A\right]
\right\rangle *1. 
\]
Apply Sibner's mean-value formula (Lemma 2) to the left-hand side of the
above identity. Take 
\[
\mu =F_A;\,\tau =d\varphi ;\,\xi =x;\,\eta =0. 
\]
We obtain the inequality 
\[
\int_{B^{\prime }}\left| d(A-\varphi )\right| ^2*1\leq C(\int_{B^{\prime
}}(|F_A|+|d\varphi |)|x|*1+ 
\]
\[
\int_{B^{\prime }}\left| A-\varphi \right| \left| A\right| |\rho (Q)|\left|
F_A\right| *1+\int_{B^{\prime }}\left| d(A-\varphi )\right| \left| A\right|
^2*1) 
\]
\begin{equation}
\equiv C(i_1+i_2+i_3).
\end{equation}
We estimate the terms $i_1$, $i_2$, and $i_3$ individually. If $R$ is the
radius of $B^{\prime },$ 
\begin{equation}
i_1=\int_{B^{\prime }}(|F_A|+|d\varphi |)|x|*1\leq C\left( \left\|
F_A\right\| _\infty +\left\| d\varphi \right\| _\infty \right)
\int_0^R|x|^nd|x|=CR^{n+1}.
\end{equation}
In an exponential gauge we have additionally, for $R<1,$ 
\[
i_2=\int_{B^{\prime }}\left| A-\varphi \right| \left| A\right| |\rho
(Q)|\left| F_A\right| *1\leq 
\]
\[
C(\rho )\left( \int_{B^{\prime }}\left| A\right| ^2\left| F_A\right|
*1+\int_{B^{\prime }}\left| \varphi \right| \left| A\right| \left|
F_A\right| *1\right) 
\]
\begin{equation}
\leq C\left( \rho ,\left\| F_A\right\| _\infty ,\left| \varphi \right|
_{C^{0,\gamma }}\right) \left[ \int_0^R\left( |x|^{n+1}+|x|^n\right)
d|x|\right] \leq CR^{n+1},
\end{equation}
where $\gamma $ is the H\"{o}lder exponent of $\varphi .$ Young's inequality
implies that there is a small positive number $\varepsilon $ for which 
\[
i_3=\int_{B^{\prime }}\left| d(A-\varphi )\right| \left| A\right| ^2*1\leq 
\]
\[
\varepsilon \int_{B^{\prime }}\left| d(A-\varphi )\right| ^2*1+C(\varepsilon
)\int_{B^{\prime }}\left| A\right| ^4*1\leq 
\]
\[
\varepsilon \int_{B^{\prime }}\left| d(A-\varphi )\right| ^2*1+C\left(
\varepsilon ,\left\| F_A\right\| _\infty \right) \int_0^R\left| x\right|
^{n+3}d|x| 
\]
\begin{equation}
=\varepsilon \int_{B^{\prime }}\left| d(A-\varphi )\right| ^2*1+CR^{n+4}.
\end{equation}
Substituting inequalities (20)-(22) into inequality (19) and absorbing small
terms on the left, we obtain 
\begin{equation}
\int_{B^{\prime }}\left| d(A-\varphi )\right| ^2*1\leq CR^{n+1}.
\end{equation}
Now we use the fact that mean value minimizes variance over all location
parameters. This allows us to replace (23) by the inequality 
\[
\int_{B^{\prime }}\left| dA-(dA)_{R,0}\right| ^2*1\leq CR^{n+1}, 
\]
where $(f)_{r,\sigma }$ denotes the mean value of $f$ in an \textit{n}-disc
of radius $r$ centered at the point $\sigma \in \mathbf{R}^n.$ Thus $dA$ is
H\"{o}lder continuous on $B^{\prime },$ with H\"{o}lder exponent 1/2, by
Campanato's Theorem (\textit{c.f.} [\textbf{Gi}], Ch.3).

Regarding the H\"{o}lder continuity of the curvature, we have, using the
linearity of the mean-value operator over sums, 
\[
\int_{B^{\prime }}\left| F-(F)_{R,0}\right| ^2*1=\int_{B^{\prime }}\left|
dA+A\wedge A-(F)_{R,0}\right| ^2*1\leq 
\]
\[
C\left( \int_{B^{\prime }}\left| dA-(F)_{R,0}\right| ^2*1+\int_{B^{\prime
}}\left| A\right| ^4*1\right) \leq 
\]
\[
C\left( \int_{B^{\prime }}\left| dA-\left[ (dA)_{R,0}+\left( A\wedge
A\right) _{R,0}\right] \right| ^2*1+R^{n+4}\right) \leq 
\]
\[
C\left( \int_{B^{\prime }}\left| dA-(dA)_{R,0}\right| ^2*1+\int_{B^{\prime
}}\left| \left( A\wedge A\right) _{R,0}\right| ^2*1+R^{n+4}\right) \leq 
\]
\[
C\left( R^{n+1}+R^{n+4}+\int_{B^{\prime }}\left| \frac 1{\left| B^{\prime
}\right| }\int_{B^{\prime }}A\wedge A*1\right| ^2*1\right) \leq 
\]
\[
C\left[ R^{n+1}+\frac 1{R^{2n}}\int_0^R\left( \left\| F\right\| _\infty
^2\int_0^R\left| x\right| ^{n+1}d|x|\right) ^2\left| x\right|
^{n-1}d|x|\right] 
\]
\begin{equation}
\leq C\left( R^{n+1}+R^{n+4}\right) \leq CR^{n+1}.
\end{equation}
Thus $F$ is H\"{o}lder continuous in $B^{\prime }$ by Campanato's Theorem.

We would like to finish the proof of Theorem 1 by covering $\Omega $ with
small \textit{n}-discs and repeating the above argument in each disc. We
would \textit{like} to do this but cannot yet. The obstacle is our use of
the exponential gauge \textit{at the origin of coordinates. }We must show
that the Campanato estimate (24) is invariant under continuous gauge
transformations in a small ball. Precisely, we show that if $F$ satisfies
(24) and if a map $\gamma \in AutX$ is continuous at each point $x\in
B_r(\sigma ),$ where $B$ is an \textit{n}-disc of sufficiently small radius $%
r$ centered at a point $\sigma $ sufficiently close to the origin, we have 
\[
\left\| \gamma ^{-1}(x)F(x)\gamma (x)-\left[ \gamma ^{-1}(x)F(x)\gamma
(x)\right] _{r,\sigma }\right\| _2\leq Cr^\beta 
\]
for $\beta >0,$ where $\left\| \cdot \right\| _2$ is now the $L^2$-norm on $%
B_r(\sigma ).$ Using the continuity of $\gamma $ to approximate the term $%
\left[ \gamma ^{-1}(x)F(x)\gamma (x)\right] _{r,\sigma }$ by $\left[ \gamma
^{-1}(\sigma )F(x)\gamma (\sigma )\right] _{r,\sigma }$ for small $r$ and
also using the fact that $\gamma $ is unitary, we have 
\[
\left\| \gamma ^{-1}(x)F(x)\gamma (x)-\left[ \gamma ^{-1}(x)F(x)\gamma
(x)\right] _{r,\sigma }\right\| _2\approx 
\]
\[
\left\| \gamma ^{-1}(x)F(x)\gamma (x)-\left[ \gamma ^{-1}(\sigma )F(x)\gamma
(\sigma )\right] _{r,\sigma }\right\| _2= 
\]
\[
\left\| F(x)-\gamma (x)\gamma ^{-1}(\sigma )\left[ F(x)\right] _{r,\sigma
}\gamma (\sigma )\gamma ^{-1}(x)\right\| _2= 
\]
\[
\left\| F(x)\gamma (x)\gamma ^{-1}(\sigma )-\gamma (x)\gamma ^{-1}(\sigma
)\left[ F(x)\right] _{r,\sigma }\right\| _2\leq 
\]
\[
\left\| F(x)\left( \gamma (x)\gamma ^{-1}(\sigma )-I\right) +\left( I-\gamma
(x)\gamma ^{-1}(\sigma )\right) \left[ F(x)\right] _{r,\sigma }\right\|
_2+\left\| F(x)-\left[ F(x)\right] _{r,\sigma }\right\| _2, 
\]
where $I$ is the identity transformation. But this is equivalent to 
\[
\left\| \gamma ^{-1}(x)F(x)\gamma (x)-\left[ \gamma ^{-1}(x)F(x)\gamma
(x)\right] _{r,\sigma }\right\| _2\leq 
\]
\begin{equation}
\left\| \left( \gamma (x)\gamma ^{-1}(\sigma )-I\right) \left( F(x)-\left[
F(x)\right] _{r,\sigma }\right) \right\| _2+Cr^{n+1}\leq C^{\prime }r^{n+1},
\end{equation}
where the inequality on the far right follows, for sufficiently small $r$
and $\sigma ,$ from (24) and the boundedness of $\gamma (x)\gamma
^{-1}(\sigma )-I.$ Inequality (25) shows that the Campanato estimate is
preserved under continuous gauge transformations in a small \textit{n}-disc
centered at a point close to the origin. Thus in applying our covering
argument we can gauge transform out of the exponential gauge and ''fan out''
from the origin, applying Campanato's Theorem in each ball as we go. Because 
$\Omega $ is a bounded type-\textit{A} domain, we will eventually cover the
entire set. This completes the proof of Theorem 1.$\Box $%
\[
\]

The following is a revision of [\textbf{O3}], Theorem 4.2. (Lemma 3 is used
as the fundamental elliptic estimate; implied conditions on dimension and
conformal weight are stated explicitly; several elliptic inequalities stated
in [\textbf{O3}] are proven; an error in the statement of a lemma in [%
\textbf{O3}] is corrected.)

\begin{theorem}
Let the pair $(A,F_A)$ smoothly satisfy eqs. (8), (9) in $B/\Sigma ,$ where $%
B$ is a small euclidean $n$-disc for $n\geq 6$ and $\Sigma $ is a Lipschitz
manifold of codimension exceeding $2n/(n-4).$ Let the section $\rho $
satisfy condition (2) with $q=0.$ If $A$ lies in the space $H^{1,n/2}(B)$
and $F_A$ lies in $L^{n/2}(B)$, then there is a continuous gauge
transformation $g$ such that the pair $(g(A),F_{g(A)})$ is a H\"{o}lder
continuous solution of (8), (9) in $B$.
\end{theorem}

\textbf{Remarks.}\textit{\ i)} We require $n\geq 6$ in order to have $%
2n/(n-4)\leq n;$ this is necessary in order for the word ''codimension'' to
make sense.

\textit{ii)} Obviously, the hypothesis on $A$ is gauge-dependent. This is
why one would call Theorem 4 a \textit{gauge-improvement theorem} rather
than a true \textit{removable singularities theorem}.

\textit{iii)} We take $\rho $ to be a section of a line bundle having
conformal weight $w$ (which may be zero) in the sense of, \textit{e.g.,} [%
\textbf{O1}].

The proof of Theorem 4 depends crucially on Theorem 1 and on the following
lemma, which extends Lemma 2.1 of [\textbf{Si2}]. (See also Theorem 3.2 of [%
\textbf{GS}].)

\begin{lemma}
(\textit{c.f.} [\textbf{O3}], Lemma 3.2). Let the $p$-form $u$ smoothly
satisfy the inequality 
\begin{equation}
-\frac \partial {\partial x^j}\left( a^{ij}(u)\frac{\partial Q}{\partial x^i}%
\right) +b^j\frac{\partial Q}{\partial x^j}-zQ\leq 0
\end{equation}
on $B/\Sigma ,$ with $a^{ij}$ satisfying the ellipticity condition $m_1|\xi
|^2\leq a^{ij}\xi _i\xi _j\leq m_2|\xi |^2,$ where $m_1$ and $m_2$ are
positive constants; $B$ is a small euclidean $n$-disc, $n>2(p+1),$ of radius 
$\tau ,$ centered at the origin of coordinates in $\mathbf{R}^n$; $\Sigma $
is a compact singular set, completely contained in $B$, of codimension $\mu $%
, where $n\geq \mu >2n/(n-2p);Q=*(u\wedge *u).$ Assume that the $L^{n/2}$%
-norm of $z$ is sufficiently small on a small ball $B^{\prime }\subset
\subset $ $B$ and that the functions $b^j$ are all in $L^s(B)$ for some $s>n$%
. If $u\in L^{n/p}(B),$ then $u\in L^s(B)\forall s<\infty .$
\end{lemma}

\textbf{Remark.} I regret that in the course of a prepublication reduction
in the length of [\textbf{O3}] I rather garbled the statement of Lemma 3.2.
In particular, the lemma concludes that $u$ is in $L^\infty (B).$ That
conclusion is unwarranted unless $z$ lies in some higher $L^P$ space than $%
L^{n/2}$ or is smoothed by the conclusion of the lemma. As it happens, in
each of the two applications of Lemma 3.2 in [\textbf{O3}], one of these
ameliorating conditions is satisfied and the conclusion of the lemma is in
fact justified in the two special cases in which it is used. Lemma 5 gives a
corrected statement of [\textbf{O3}], Lemma 3.2 . (A correct version was
also circulated as a prepublication preprint of [\textbf{O3}].)

\textit{Proof of Lemma 5.} Integrate inequality (26) against a nonnegative
test function $\zeta \in C_0^\infty (B)$ which vanishes on $\Sigma .$
Precisely, let 
\[
\zeta =\left( \eta \psi \right) ^2\Xi (Q), 
\]
where $\eta ,\psi \geq 0;\psi (x)=0\,\forall x$ in a neighborhood of $\Sigma
;\eta \in C_0^\infty (B^{\prime })$ where $B^{\prime }\subset \subset B$ is
chosen small enough so that the $L^{n/2}$ norm of $z,$ and the $L^n$ norm of 
$b\equiv \left| \sum_jb^j\right| ,$ are small on $B^{\prime }$ ; $\Xi
(Q)=H(Q)H^{\prime }(Q),$ where $H(Q)=H_\kappa (Q)$ is the following variant
of Serrin's test function [\textbf{Se1}]: 
\[
H_\kappa (Q)= \left\{ 
\begin{array}{l}
Q^{[n/(n-2)]^\kappa n/4p}\;for\;0\leq Q\leq \ell , \\ 
\frac{\mu -\varepsilon }{\mu -2-\varepsilon }\left[ \left( \ell \cdot
Q^{(\mu -2-\varepsilon )/2}\right) ^{[n/(n-2)]^\kappa n/2p(\mu -\varepsilon
)}-\frac 2{\mu -\varepsilon }\ell ^{[n/(n-2)]^\kappa n/4p}\right] for\;Q\geq
\ell .
\end{array}
\right. 
\]
Notice that $H_\kappa (Q)$ is finite $\forall \kappa <\infty $ but that $%
H_\kappa (Q)$ is singular if $\kappa $ is infinite. Iterate a sequence of
elliptic estimates, taking successively $u\in L^{\alpha (\kappa )}(B)$ for $%
\alpha (\kappa )=[n/(n-2)]^\kappa (n/p),\kappa =0,1,...\,.$ For each $\kappa 
$ we have 
\[
\int_{B^{\prime }}a^{ij}(u)\partial _iQ\cdot 2\left( \eta \psi \right)
\partial _j\left( \eta \psi \right) \Xi (Q)*1+\int_{B^{\prime
}}a^{ij}(u)\left( \eta \psi \right) ^2\Xi ^{\prime }(Q)\partial _iQ\partial
_jQ*1 
\]
\[
\leq \int_{B^{\prime }}|z|Q\left( \eta \psi \right) ^2\Xi
(Q)*1+\int_{B^{\prime }}|b|\left| \nabla Q\right| \left( \eta \psi \right)
^2\Xi (Q)*1. 
\]
This inequality can be rewritten in the short-hand form 
\[
I_1+I_2\leq I_3+I_4 
\]
or more conveniently 
\begin{equation}
I_2\leq I_3+I_4+|I_1|,
\end{equation}
the integrals of which we estimate individually.

The definitions of $\Xi $ and $H$ imply the inequalities 
\begin{equation}
\Xi ^{\prime }(Q)\geq C\left( H^{\prime }(Q)\right) ^2
\end{equation}
and 
\begin{equation}
Q\Xi \leq \left( \frac n{n-2}\right) ^\kappa \frac n{4p}H^2.
\end{equation}
(A wish to satisfy (28) is the motivation behind the lower bound on $\mu .)$
Inequality (28) and the ellipticity condition imply that 
\[
I_2=\int_{B^{\prime }}a^{ij}(u)\left( \eta \psi \right) ^2\Xi ^{\prime
}(Q)\partial _iQ\partial _jQ*1\geq 
\]
\begin{equation}
C(m_1)\int_{B^{\prime }}\left( \eta \psi \right) ^2\left( H^{\prime
}(Q)\right) ^2\left| \nabla Q\right| ^2*1=C\int_{B^{\prime }}\left( \eta
\psi \right) ^2\left| \nabla H\right| ^2*1.
\end{equation}
We have by Young's inequality 
\[
I_1=\int_{B^{\prime }}a^{ij}(u)\partial _iQ\cdot 2\left( \eta \psi \right)
\partial _j\left( \eta \psi \right) H(Q)H^{\prime }(Q)*1= 
\]
\[
2\int_{B^{\prime }}\left( a^{ij}(u)\left( \eta \psi \right) \left( \partial
_iH\right) \right) \partial _j\left( \eta \psi \right) H*1\leq 
\]
\[
m_2\left( \varepsilon \int_{B^{\prime }}\left( \eta \psi \right) ^2\left|
\nabla H\right| *1+C(\varepsilon )\int_{B^{\prime }}\left| \nabla \left(
\eta \psi \right) \right| ^2H^2*1\right) \equiv i_{11}+i_{12}. 
\]
For small $\varepsilon ,$ the integral $i_{11}$ can be subtracted from the
lower bound on $I_2$ in (30). Using inequality (29) and the Sobolev
inequality, we can write 
\[
I_3=\int_{B^{\prime }}|z|Q\left( \eta \psi \right) ^2\Xi (Q)*1\leq \left(
\frac n{n-2}\right) ^\kappa \frac n{4p}\int_{B^{\prime }}|z|\left( \eta \psi
\right) ^2H^2*1\leq 
\]
\[
C\left\| z\right\| _{n/2}\left( \int_{B^{\prime }}\left( \eta \psi H\right)
^{2n/(n-2)}*1\right) ^{(n-2)/n}\leq C^{\prime }\left\| z\right\|
_{n/2}\left\| \eta \psi H\right\| _{1,2}^2. 
\]
Expanding the term on the right by the product rule and using Young's
inequality, we have 
\[
I_3\leq C\left\| z\right\| _{n/2}\left\{ \int_{B^{\prime }}\left[ \left|
\nabla \left( \eta \psi \right) \right| ^2+\left( \eta \psi \right)
^2\right] H^2*1+\int_{B^{\prime }}\left( \eta \psi \right) ^2\left| \nabla
H\right| ^2*1\right\} 
\]
\begin{equation}
\equiv i_{31}+i_{32}.
\end{equation}
The integral $i_{32}$ can be subtracted from the lower bound on $I_2$ in
(30), as $B^{\prime }$ has been chosen so that the product of our
independent constant $C$ and the $L^{n/2}$ norm of $z$ is small. Young's
inequality yields 
\[
I_4=\int_{B^{\prime }}|b|\left| \nabla Q\right| \left( \eta \psi \right)
^2H(Q)H^{\prime }(Q)*1=\int_{B^{\prime }}|b|\left( \eta \psi \right)
^2H\left| \nabla H\right| *1\leq 
\]
\begin{equation}
C(\varepsilon )\int_{B^{\prime }}|b|^2\left( \eta \psi \right)
^2H^2*1+\varepsilon \int_{B^{\prime }}\left( \eta \psi \right) ^2\left|
\nabla H\right| ^2*1.
\end{equation}
We can write inequality (32) in the short-hand form 
\[
I_4\leq i_{41}+i_{42}. 
\]
We similarly rewrite the integral inequality (27) in the form 
\begin{equation}
I_2-\left( i_{32}+i_{42}+i_{11}\right) \leq C\left(
i_{12}+i_{31}+i_{41}\right) .
\end{equation}
Notice that the left-hand side of (33) remains nonnegative for small $%
\varepsilon $ when $I_2$ is replaced by the extreme right-hand integral in
(30). Moreover, 
\begin{equation}
i_{41}\leq C\left\| b\right\| _n^2\left\| \eta \psi H\right\|
_{2n/(n-2)}^2\leq C\left\| b\right\| _n^2\left\| \eta \psi H\right\|
_{1,2}^2,
\end{equation}
which is analogous to $I_3$. Because $b\in L^s$ for $s>n,$ the $L^n$-norm of 
$b$ is small on $B^{\prime }$ if $B^{\prime }$ is small. We simultaneously
estimate the terms of $i_{31}$ and $i_{41}$ which involve the gradient of $%
\psi .$ There exists ([\textbf{Se2}], \textit{c.f.} Lemma 2 and p. 73) a
sequence of functions $\xi _\nu $ such that:

\textit{a)} $\xi _\nu \in [0,1]\;\forall \nu ;$

\textit{b)} $\xi _\nu \equiv 1$ in a neighborhood of $\Sigma \;\forall \nu ;$

\textit{c)} $\xi _\nu \rightarrow 0$ a.e. as $\nu \rightarrow \infty ;$

\textit{d)} $\nabla \xi _\nu \rightarrow 0$ in $L^{\mu -\varepsilon }$ as $%
\nu \rightarrow \infty .$

Apply the product rule to the squared $H^{1,2}$ norms in $i_{31}$
[inequality (31)] and in $i_{41}$ [inequality (34)] letting $\psi =\psi _\nu
=1-$ $\xi _\nu .$ Observing that the cross terms in $\left( \nabla \eta
\right) \psi $ and $\left( \nabla \psi \right) \eta $ can be absorbed into
the other terms by applying Young's inequality, we estimate 
\[
\lim_{\nu \rightarrow \infty }\int_{B^{\prime }}\eta ^2\left| \nabla \psi
_\nu \right| ^2H^2*1\leq \lim_{\nu \rightarrow \infty }C(\ell
)\int_{B^{\prime }}\left| \nabla \psi _\nu \right| ^2Q^{\frac{\mu
-2-\varepsilon }{\mu -\varepsilon }\left( \frac n{n-2}\right) ^\kappa \frac
n{2p}}*1 
\]
\begin{equation}
\leq \lim_{\nu \rightarrow \infty }C(\ell )\left\| \nabla \psi _\nu \right\|
_{\mu -\varepsilon }^2\left\| u\right\| _{\alpha (\kappa )}^{\alpha (\kappa
)(\mu -2-\varepsilon )/(\mu -\varepsilon )}=0.
\end{equation}
Having shown that the integral on the left in (35) is zero for every value
of $\ell ,$ we can now let $\ell $ tend to infinity. We obtain via Fatou's
Lemma the inequality 
\begin{equation}
\int_{B^{\prime }}\eta ^2\left| \nabla \left( Q^{\alpha (\kappa )/4}\right)
\right| ^2*1\leq \int_{B^{\prime }}\left| \nabla \eta \right| ^2Q^{\alpha
(\kappa )/2}*1.
\end{equation}
Thus $Q^{\alpha (\kappa )/4}$ is in $H^{1,2}$ on some smaller disc on which $%
\eta =1.$ But then, because $u$ is assumed to be smooth away from the
singularity and $\Sigma $ is compact, $Q^{\alpha (\kappa )/4}$ must be in $%
H^{1,2}$ on the larger disc as well. Now apply the Sobolev inequality to
conclude that $u$ is now in the space $L^{\alpha (\kappa +1)}(B).$ Because
the sequence $\left\{ n/(n-2)\right\} ^\kappa $ obviously diverges, we
conclude after a finite number of iterations of this argument that $Q^c$ is
in $H^{1,2}(B)$ for any positive value of $c.$ A final application of the
Sobolev inequality yields the assertion of Lemma 5.$\Box $

\textit{Proof of Theorem 4.} Use (15) to apply Lemma 5 with $u=|F|$, $%
z=|\nabla A|+|A|^2,$ and $b^j\equiv 0\,\forall j.$ For $F\in L^{n/2}$ the
small-norm assumptions of the lemma follow from conventional scaling
arguments in a Hodge gauge (see the comments following Theorem 1.3 of [%
\textbf{U3}]). Remark \textit{iii)} following the statement of Theorem 4
implies that eqs. (8), (9) remain unaffected by such changes of scale. We
conclude that $F$ is in $L^p$ for some $p>n/2.$ The continuous gauge
transformation guaranteed by Theorem 1.3 of [\textbf{U3}] can be applied up
to the boundary of $\Sigma ,$ using the methods of [\textbf{SS5}], as $%
\Sigma $ is Lipschitz. Because the test functions used in proving Lemma 5
did not require any more smoothness than is implied by the definition of
weak solution adopted in Remark \textit{ii)} following Theorem 1, we can use
(36) to show that $F$ is a weak solution of (8), (9) in all of $B$. Theorem
1 now implies that $F$ is H\"{o}lder continuous in $B$.$\Box $

\textbf{Remark.} A similar argument implies a removable singularities
theorem for solutions of eqs. (3), (4). Apply Lemma 5, taking $z$ to be an
upper bound on the sectional curvature of the restriction to $B$ of the
Riemannian manifold $M$. Thus $z$ can be chosen to be zero for a
sufficiently small singular set. For any singular set $z$ can be chosen to
lie in a higher $L^P$ space than $n/2$, as the singularity is in $T^{*}M$
rather than in the metric on $M$. In this case the arguments of the lemma
imply, using Theorem 5.3.1 of [\textbf{Mo}], that the $p$-form $\omega $
(taking  $u=\left| \omega \right| )$ is a bounded weak solution of eqs. (3),
(4) in $B$. Observing that the arguments of [\textbf{S1}], Section 4 are
local, we can apply them in $B$ to conclude that $\omega $ is H\"{o}lder
continuous in all of $B$. Details are given in Theorem 3.1 of [\textbf{O3}].

\section{Appendix}

In this section we sketch the proof of a technical lemma required for the
proof of Theorem 1. Details are given in [\textbf{O2}], pp. 387-392.

\begin{lemma}
([\textbf{O2}], Lemma 3.1). Under the hypotheses of the theorem, $F_A\in
H^{1,2}(B)$ for sufficiently small $n$-disc $B$.
\end{lemma}

\textit{Proof.} It is sufficient to prove Lemma 6 in a Hodge gauge. In eq.
(14) replace the admissible test function $\zeta (x)$ by the admissible test
function $\zeta (x-he_i),$ where $e_i$ is the $i^{th}$ basis vector for $%
\mathbf{R}^n$, $i=1,\,...,\,n,$ and $h$ is a positive constant. Then eq.
(14) assumes the form 
\begin{equation}
\int_B\left\langle d\zeta (x-he_i),\rho \left( Q(x)\right) F(x)\right\rangle
d^nx=-\int_B\left\langle \zeta (x-he_i),*\left[ A(x),*\rho \left(
Q(x)\right) F(x)\right] \right\rangle d^nx.
\end{equation}
If we subject both sides of identity (37) to the coordinate transformation $%
y=x-he_i,$ subtract eq. (14) from the resulting equation and divide through
by $h$, we obtain 
\[
\int_B\left\langle d\zeta (x),*\frac{\left\{ \rho \left( Q(x+he_i)\right)
F(x+he_i)-\rho \left( Q(x)\right) F(x)\right\} }h\right\rangle d^nx= 
\]
\[
-\int_B\left\langle \zeta (x),*\left[ \Delta _{i,h}A(x),*\rho \left(
Q(x+he_i)\right) F(x+he_i)\right] \right\rangle d^nx- 
\]
\begin{equation}
\int_B\left\langle \zeta (x),\frac{\left[ A(x),*\left\{ \rho \left(
Q(x+he_i)\right) F(x+he_i)-\left( Q(x)\right) F(x)\right\} \right] }%
h\right\rangle d^nx,
\end{equation}
where 
\[
\Delta _{i,h}u(x)=\frac{u(x+he_i)-u(x)}h. 
\]
Apply Lemma 2 to each of the terms 
\[
\frac{\rho \left( Q(x+he_i)\right) F(x+he_i)-\rho \left( Q(x)\right) F(x)}h 
\]
enclosed in braces on each side of (38). Choose $\xi =x+he_i,\eta =x,\mu
=F(x+he_i),$ and $\tau =F(x).$ We obtain 
\[
\int_B\left\langle d\zeta ,\alpha ^{ij}\Delta _{i,h}F+\beta
_i^je_i\right\rangle d^nx= 
\]
\[
-\int_B\left\langle \zeta (x),*\left[ \Delta _{i,h}A(x),*\rho \left(
Q(x+he_i)\right) F(x+he_i)\right] \right\rangle d^nx 
\]
\begin{equation}
-\int_B\left\langle \zeta (x),*\left[ A(x),*\left( \alpha ^{ij}\Delta
_{i,h}F+\beta _i^je_i\right) \right] \right\rangle d^nx.
\end{equation}
The function $\zeta (x)=\eta (x)\Delta _{i,h}A(x)$ is an admissible test
function for all $h$ by [\textbf{U3}], Theorem 1.3 \textit{ii)}, provided $%
\eta \in C_0^\infty (B).$ Choose $\eta (x)$ to be nonnegative $\forall x\in
B $ and $\eta (x)\equiv 1$ for $x\in B^{\prime },$ where $B^{\prime }$ is a
ball completely contained in $B$. Using the fact that $d$ commutes with $%
\Delta _{i,h},$ we can write (39) in the form 
\[
\int_B\eta \left| \Delta _{i,h}\left( dA\right) \right| ^2d^nx\leq 
\]
\[
C\{\int_B\left( \left| A(x+he_i)\right| +\left| A(x)\right| \right) \left|
\Delta _{i,h}A\right| \left| \Delta _{i,h}\left( dA\right) \right| d^nx+ 
\]
\[
\int_B\left( \left| F(x+he_i)\right| +\left| F(x)\right| \right) \left|
\Delta _{i,h}\left( dA\right) \right| d^nx+\int_B\left| \Delta
_{i,h}A\right| ^2\left( \left| F(x+he_i)\right| \right) d^nx 
\]
\[
+\int_B\left( \left| A(x)\right| +1\right) \left| \Delta _{i,h}\left(
dA\right) \right| \left| \Delta _{i,h}A\right| d^nx+ 
\]
\[
\int_B\left( \left| A(x)+1\right| \right) \left| \Delta _{i,h}A\right|
^2\left( \left| A(x+he_i)\right| +\left| A(x)\right| \right) d^nx 
\]
\begin{equation}
+\int_B\left( \left| F(x+he_i)\right| +\left| F(x)\right| \right) \left|
\Delta _{i,h}A\right| d^nx\}.
\end{equation}
In order to estimate the right-hand side of (40) we use the relation 
\[
\int_B\left| \nabla \left| \Delta _{i,h}A\right| \right| ^2d^nx\leq
\int_B\left| \nabla \left( \Delta _{i,h}A\right) \right| ^2d^nx 
\]
\begin{equation}
\leq C\left( \int_B\left| \Delta _{i,h}\left( dA\right) \right|
^2d^nx+\int_B\left| \Delta _{i,h}A\right| ^2d^nx\right) ,
\end{equation}
which follows from the Kato and Gaffney-G\aa rding inequalities, and from
the facts that the exterior derivative and its adjoint commutes with $\Delta
_{i,h}$ and $\delta A=0.$ Now we have 
\[
\int_B\left( \left| A(x+he_i)\right| +\left| A(x)\right| \right) \left|
\Delta _{i,h}A\right| \left| \Delta _{i,h}\left( dA\right) \right| d^nx\leq 
\]
\[
C\left( \left\| A(x+he_i)\right\| _n^2+\left\| A(x)\right\| _n^2\right)
\left( \int_B\left| \nabla \left| \Delta _{i,h}A\right| \right|
^2d^nx+\int_B\left| \Delta _{i,h}A\right| ^2d^nx\right) 
\]
\[
+\varepsilon \int_B\left| \Delta _{i,h}A\right| ^2d^nx. 
\]
Using (41) and the fact that $A$ is small in $L^n$ on $B$, terms involving $%
\left| \Delta _{i,h}\left( dA\right) \right| ^2$ can be subtracted from the
left-hand side of (40). The other terms on the right in (40) can be handled
similarly, using the fact that $F$ is small in $L^{n/2}$ on $B.$ Letting $h$
tend to zero, we obtain 
\[
\int_{B^{\prime }}\left| \nabla \left( dA\right) \right| ^2*1<\infty . 
\]
The proof of Lemma 6 is completed by applying the Sobolev inequality to $dA$
and writing 
\[
\int_{B^{\prime }}\left| \nabla F\right| ^2*1\leq C\left( \int_{B^{\prime
}}\left| \nabla \left( dA\right) \right| ^2*1+\left\| A\right\| _n^2\left\|
\nabla A\right\| _{2n/(n-2)}^2\right) .\;\Box 
\]

\[
\mathbf{References} 
\]

\textbf{[B] }L. Bers, Mathematical Aspects of Subsonic and Transonic Gas
Dynamics, Wiley, New York, 1958.

[\textbf{Ga}] M. P. Gaffney, The harmonic operator for exterior differential
forms, \textit{Proc. N. A. S}. \textbf{37} (1954), 48-50.

[\textbf{Gi]} M. Giaquinta, Multiple Integrals in the Calculus of Variations
and Nonlinear Elliptic Systems, Princeton University Press, Princeton, 1983.

[\textbf{GS}] B. Gidas and G. Spruck, Global and local behavior of positive
solutions of nonlinear elliptic equations, \textit{Commun. Pure Appl. Math.} 
\textbf{34}(1981), 525-598.

[\textbf{Ma}] A. Marini, Dirichlet and Neumann boundary value problems for
Yang-Mills connections, Ph.D. dissertation, The University of Chicago, 1990.

[\textbf{MM}] K. B. Marathe and G. Martucci, The Mathematical Foundations of
Gauge Theories, North-Holland, Amsterdam, 1992.

[\textbf{Mo}] C. B. Morrey, Jr., Multiple Integrals in the Calculus of
Variations, Springer, New York, 1966.

[\textbf{O1}] T. H. Otway, The coupled Yang-Mills-Dirac equations for
differential forms, \textit{Pacific J. Math.} \textbf{146}, No. 1 (1990),
103-113.

\textbf{[O2]} T. H. Otway, Yang-Mills fields with nonquadratic energy, 
\textit{J. Geometry \& Physics} \textbf{19 }(1996),379-398.

\textbf{[O3]} T. H. Otway, Properties of nonlinear Hodge fields, \textit{J.
Geometry \& Physics }\textbf{27}(1998), 65-78.

[\textbf{O4}] T. H. Otway, An elliptic inequality for nonlinear Hodge
fields, Los Alamos National Laboratory automated electronic archive at
xxx.lanl.gov, math-ph/9806007 (1998).

[\textbf{Se1}] J. Serrin, Local behavior of solutions of quasilinear
equations, \textit{Acta Math.} \textbf{111} (1964), 247-302.

[\textbf{Se2}] J. Serrin, Removable singularities of solutions of elliptic
equations, \textit{Archs. Ration. Mech. Analysis} \textbf{17 }(1964), 67-78.

\textbf{[Si1]} L. M. Sibner, An existence theorem for a nonregular
variational problem, \textit{Manuscripta} \textit{Math.} \textbf{43 }(1983),
45-72.

[\textbf{Si2}] L. M. Sibner, The isolated point singularity problem for the
coupled Yang-Mills equations in higher dimensions, \textit{Math. Ann.} 
\textbf{271} (1985), 125-131.

\textbf{[SS1]} L. M. Sibner and R. J. Sibner, A nonlinear Hodge-de Rham
theorem, \textit{Acta Math.} \textbf{125 }(1970), 57-73.

\textbf{[SS2]} L. M. Sibner and R. J. Sibner, Nonlinear Hodge theory:
Applications, \textit{Advances in Math.} \textbf{31} (1979), 1-15.

\textbf{[SS3]} L. M. Sibner and R. J. Sibner, A maximum principle for
compressible flow on a surface, \textit{Proc. Amer. Math. Soc.} \textbf{71}%
(1) (1978), 103-108.

\textbf{[SS4]} L. M. Sibner and R. J. Sibner, A subelliptic estimate for a
class of invariantly defined elliptic systems, \textit{Pacific J. Math.} 
\textbf{94}(2) (1981), 417-421.

[\textbf{SS5}] L. M. Sibner and R. J. Sibner, Classification of singular
Sobolev connections by their holonomy, \textit{Commun. Math. Phys. }\textbf{%
144}(1992), 337-350.

\textbf{[T]} Tchrakian, D. H., N-dimensional instantons and monopoles, 
\textit{J. Math. Physics} \textbf{21} (1980), 166-169.

\textbf{[U1]} K. Uhlenbeck, Regularity for a class of nonlinear elliptic
systems, \textit{Acta Math.} \textbf{138 }(1977), 219-240.

\textbf{[U2] }K. Uhlenbeck, Removable singularities in Yang-Mills fields, 
\textit{Commun. Math. Physics} \textbf{83 }(1982), 11-30.

\textbf{[U3]} K. Uhlenbeck, Connections with $L^p$ bounds on curvature,%
\textit{\ Commun. Math. Physics} \textbf{83 }(1982), 31-42.

\end{document}